\documentstyle[psfig,amssymb]{aa}
\newcommand{\be}{\begin{equation}}
\newcommand{\ee}{\end{equation}}
\begin{document}


\title{X-ray absorption lines suggest matter infalling onto the central 
black-hole of Mrk 509}
\author{M. Dadina,
  \inst{1}
  M. Cappi, 
  \inst{1}
  G. Malaguti, 
  \inst{1}
   G. Ponti 
  \inst{1,2,3}
   \and
  A. De Rosa, 
  \inst{4}
   }

\institute{
\inst{1}IASF/INAF Sezione di Bologna, via Gobetti 101, I-40129 Bologna, Italy.\\
\inst{2}Dipartimento di Astronomia dell'Universit\'a degli Studi di Bologna, via Ranzani 1, I-40127, Bologna, Italy.\\
\inst{3}Institute of Astronomy, Madingley Road, Cambridge CB3 0HA, UK\\  
\inst{4}IASF/INAF, via Fosso del Cavaliere 100, I-00133 Roma, Italy.\\
\email{dadina@bo.iasf.cnr.it}
}

\date{Received date/ Accepted date}

\abstract{

Evidence for  both red- and blue-shifted absorption lines 
due to ionized Fe in the X-ray spectrum of the Seyfert 1 galaxy Mrk 509 
are reported. 
These features appear to be transient on time-scales as short 
as $\sim$20 ks, 
and have been observed with two different satellites, $BeppoSAX$ 
and {\it XMM-Newton}. 
The red- and blue-shifted lines are found at E$\sim$5.5 keV and 
$\sim$8.1-8.3 keV (rest-frame), respectively. The first is seen in one out of six $BeppoSAX$ observations, the latter is seen by both satellites. 
Under the assumption that the absorption is due to either H- or He-like 
Iron, the implied velocities for the absorbing matter are v$\sim$0.15-0.2 c,
in both outward and inward directions.
An alternative explanation in terms of gravitational red-shift 
for the $\sim$5.5 keV line cannot be ruled out with the current data. 
We argue, however, that the 
temporal patterns and sporadic nature of the lines are more easily reconciled 
with models which predict important radial motions close to the central black 
hole, such as the ``aborted jet'' model, the ``thundercloud'' model, or 
magneto-hydrodynamical models of jets and accretion-disks.

\keywords{Galaxies: Seyfert -- X-rays: individual: Mrk 509 -- Black hole physics}
  }
\authorrunning{M. Dadina et al.}
\titlerunning{Absorption Lines in Mrk 509}
\maketitle

\section{Introduction}

Most popular models adopted to explain the high 
luminosity of AGNs are based on the release of gravitational energy of 
infalling matter accreting onto a super massive black hole (SMBH). 
Nonetheless few are
the direct measurements of matter infall/accretion in AGNs. The most 
impressive one is probably the detection of  a 
relativistically broadened FeK${\alpha}$ line in the X-ray spectrum of 
the Seyfert 1 galaxy  MCG-6-30-15
(Tanaka et al. 1995). Detailed studies of the FeK$_{\alpha}$ double-horn
profile have the potential to trace the matter path in the vicinity of the 
SMBH. However, recent results obtained with XMM-Newton and  Chandra 
observatories show that data are more complex to interpret than expected 
(Reeves et al. 2001; Bianchi et al. 2004; Dovciak et al. 2004). 
Probably, the most convincing 
evidence of matter accreting onto a SMBH are,  
to date, the one presented by Ponti et al. (2004) on MCG-6-30-15 and by 
Iwasawa, Miniutti \& Fabian (2004) on NGC 3516. Using two different
model-independent and timing tools, these authors showed the existence of a 
highly variable component at the energies of the putative relativistically 
broadened and red-shifted wing of the FeK${\alpha}$ line. 

Several are, on the contrary, the evidences of matter outflows in 
AGNs. The most spectacular displays of such matter ejecta are the 
jets observed in radio loud AGNs. Recently, absorption and 
emission line-like features due to outflowing matter have been also detected 
in the X-ray spectra of some quasars 
(Pounds et al. 2003a, Pounds et al. 2003b, 
Dadina \& Cappi 2003) and Seyfert galaxies (Kaspi et al. 2000; Kaastra et 
al. 2000; Turner, Kraemer \& Reeves 2004; Reeves, O'Brien \& Ward 2004). 

\begin{center}
\begin{table*}
\hspace{3.5cm}\hspace{3.0cm}\begin{minipage}{170mm}
\tabcolsep=3.0mm
\footnotesize
\caption{Observation Log. Column I: Name of the observatory. Column II: name used in the main text to indicate the observation. Column III: date of the observation. Column IV: overall duration of the observation; Column V: $BeppoSAX$ MECS or $XMM$-$Newton$ EPIC pn net exposure. Column VI: 2-10 keV flux.}
\scriptsize
\hspace{2.0cm}\begin{tabular}{c c c c c c}
\hline
\hline
&&&&&\\
Satellite & Name & Date& Duration & Exposure & F$_{\rm 2-10\; keV}$ \\
&&&&&\\
&&&(ks)&(ks)&(10$^{-11}$ erg s$^{-1}$ cm$^{-2}$)\\
&&&&&\\
I&II&III&IV&V&VI\\
&&&&&\\
\hline
&&&&&\\
BeppoSAX &    & May 18, 1998 &  110 &    52    & 5.0  \\
&&&&&\\
BeppoSAX &   & October 11, 1998 & 79  & 36  &  6.2  \\
&&&&&\\
XMM-Newton&XMM1 & October 25, 2000  & 31 & 21 & 2.6  \\
&&&&&\\
BeppoSAX &   & November 3, 2000 &  87 & 41 & 2.9  \\
&&&&&\\
BeppoSAX &  & November 8, 2000 &  85 & 38 & 2.5  \\
&&&&&\\
BeppoSAX &  & November 18, 2000 & 93 & 40 & 2.5 \\
&&&&&\\
BeppoSAX&SAX6   & November 24, 2000 & 91 & 33 &  2.6 \\
&&&&&\\
XMM-Newton&  & April 20, 2001 &  44 & 23 & 4.0 \\
&&&&&\\
\hline
\hline
\end{tabular}
\end{minipage}
\end{table*}
\end{center}

\vspace{-0.9cm}

Nandra et al. (1999) discussed the possible presence of a red-shifted Fe
absorption feature in the X-ray spectrum of the Seyfert 1 galaxy NGC 3516. 
The ASCA data showed a sharp and narrow count drop at E$\sim$5.9 keV.
Being the line red-shifted, the authors speculated that the matter
was infalling and/or suffering gravitational red-shift close to the central
black hole. Recently, further evidences of red-shifted absorption 
lines in the X-ray spectra of luminous QSOs have been claimed by 
Yaqoob \& Serlemitsos (2005) and Matt et al (2005).


Mrk 509 ($z$=0.034, Fisher et al. 1995) has been observed in X-rays several 
times with a 2-10 keV flux varying 
between  $\sim$2-5$\times$10$^{-11}$ erg cm$^{-2}$ s$^{-1}$. Its X-ray 
continuum is 
quite ``typical'' (photon index $\Gamma$$\sim$1.6-1.9,  Turner \& Pounds al. 1989, 
Nandra \& Pounds 1994) for a Seyfert 1 galaxy. 
At energies below $\sim$2 keV the spectrum is dominated by a soft excess that 
has been modeled by a warm absorber (Reynolds 1997, George et al. 1998), 
and/or an extra emission component due to ionized reflection or to the 
big blue bump hard tail (Perola et al. 2000; De Rosa et al. 2004).
$BeppoSAX$ observations indicated the presence of a reflection component, 
and of a high energy cut-off at E$\sim$70 keV (Perola et al. 2000).

A weak FeK$_{\alpha}$ emission line was firstly observed with Ginga at 
E=6.6$\pm$0.3 keV (Nandra \& Pounds 1994), and then  confirmed by ASCA 
(Nandra et al. 1997). 
This feature was, nonetheless, detected only in 5 out of the 11 ASCA 
observations (Weaver, Gelbord, \& Yaqoob 2001); in one case, 
the line was detected at E$\sim$7 keV.

Broad band ($\sim$0.1-100 keV) analysis of the six $BeppoSAX$ 
observations of Mrk 509 was performed by De Rosa et al. (2004) by merging 
the two 1998 and the four 2000 observations in two average spectra. 
These authors compared two alternative models: a cold Compton reflection and a 
reflection from an ionized disk. 
For both sets of observations, the authors reported the detection of a narrow 
and weak FeK${\alpha}$ emission line at $\sim$6.4 keV, 
with equivalent width (EW) $\sim$60 and 100 eV, for the 1998 and 2000 
observations respectively. 
The intensity of the line, as well as its constancy and narrowness lead the 
authors to argue that this component arises from distant, cold and  optically 
thick matter (De Rosa et al. 2004; Yaqoob et al. 2003), probably associated to
the optical broad line regions. 

{\it XMM-Newton} observed Mrk 509 twice. During the first observation 
(Pounds et al. 2001; Page, Davis \& Salvi 2003), a very weak (EW$\sim$50 eV) 
and narrow  
FeK${\alpha}$ line was detected at 6.4 keV. Signatures of a possible second 
line at E$\sim$6.7-6.9 keV were also observed. To account for this ionized 
line and to partially explain the soft-excess below 1.5 keV, 
the authors used a model based on the enhanced reflection due to ionized 
matter. 

\vspace{-0.3truecm}
\section{Data reduction and analysis}

We analyzed all the $BeppoSAX$ and {\it XMM-Newton} observations of Mrk 509 
(table 1) although in the next chapters
our attention is mainly focused on the 
$BeppoSAX$ observation performed on November 24, 2000 (hereafter SAX6) 
and on the {\it XMM-Newton} pointing of October 25, 2000 
(hereafter XMM1).

Source counts were extracted from circular regions 
with radius equal to 8$^{\prime}$ and 4$^{\prime}$ for $BeppoSAX$ 
LECS and MECS, respectively. 
The LECS, MECS and PDS spectra were rebinned 
so as to sample the energy resolution of the instruments using grouping files 
produced by the ASI Science Data Center. 
To subtract the background, we used 
standard PHA files accumulated from observations of empty sky regions and 
produced by the $BeppoSAX$ team. We also checked that the results obtained 
using  local backgrounds were consistent with the ones presented here. 
The LECS and MECS data were used only in the 0.1-3.5 keV and 1.7-10 keV 
energy ranges respectively, i.e. where the instruments are best calibrated.
Most recent calibration files were used for the spectral analysis.

{\it XMM-Newton} observed Mrk 509 with the EPIC CCDs operating in small window
mode. The extraction regions were circles of 35'' for both source 
and background. Latest version of SAS was used to reduce the data and
latest available calibration files were used in data analysis. Data were 
rebinned so as to have a minimum of 35 counts per bin.

\subsection {SAX6 observation: whole exposure}

Figure 1 shows the SAX6 broad band ($\sim$0.1-100 keV) spectrum 
(upper panel) and residuals (lower panel) when data are fitted with a 
power-law with Galactic absorption 
(N$_{H,gal}$$\sim$4$\times$10$^{20}$ cm$^{-2}$, Dickey \& Lockman 1990). 
As clearly visible there 
are some major spectral features, namely a soft X-ray excess below 
$\sim$1 keV, and some narrow emission and absorption features between 
$\sim$5-7 keV. 

\begin{figure}[htb]
\psfig{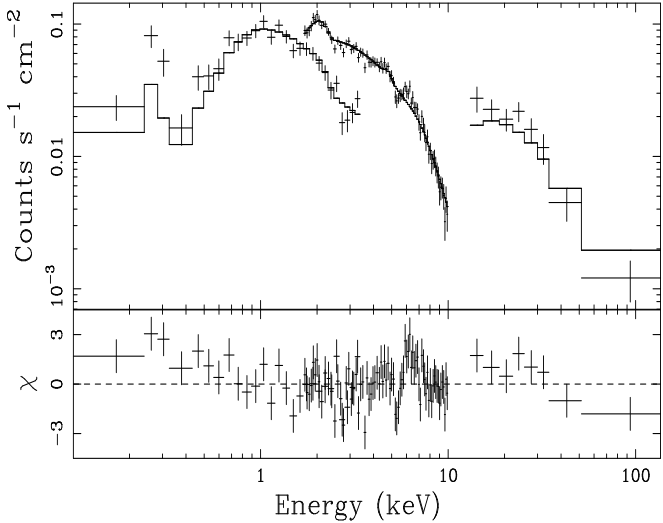}
\caption{Upper panel: broad-band spectrum of Mrk 509 during the SAX6 observation. Lower panel: data-to-model residuals (expressed in terms of standard deviations) assuming a simple power-law model with Galactic absorption.}
\end{figure}


Figure 2 (panel a) shows in greater detail the 3-10 keV data-to-model ratios 
for a simple power law model fit. 
These clearly indicate an emission feature close to the energies typical of 
the FeK$\alpha$ line and the possible presence of an absorption feature at E$\sim$5.3 keV.
To exclude that this count drop could be due to a calibration artifact of the 
response matrices and to adopt a model-independent representation of the data, 
we calculated the PHA ratios between Mrk 509 and the reference spectrum of 
3C 273 acquired on January 9, 2000.  
3C273 was chosen since it ensures a very good statistics and has an 
almost featureless spectrum with a flat photon index similar to the one of 
Mrk 509 during SAX6. We have checked that the 
FeK$\alpha$ line known to be present in the $BeppoSAX$ spectrum of 3C 273 
(Grandi \& Palumbo 2004, and references therein) is much too weak 
(less than $\sim$5\% above the continuum) to have a significant impact on 
our conclusions on the absorption feature measured between 5-6 keV.
The PHA ratios (figure 2, panel b) are indeed very similar to what shown in 
panel a) of figure 2, thus excluding response matrix problems and/or 
an erroneous modeling of the data.

\begin{figure}[htb]
\psfig{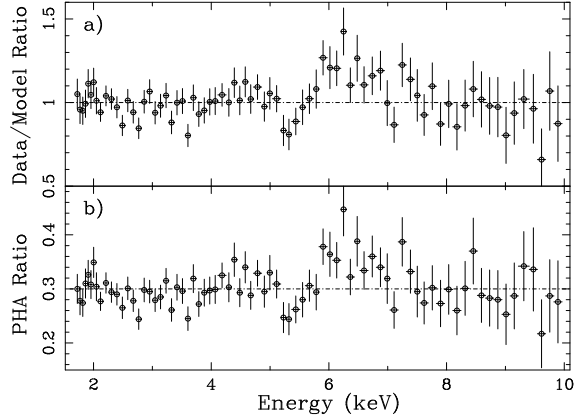}
\caption{Panel a: data-to-model ratios for SAX6 when the data
are fitted with a simple power law model. Panel b: PHA ratios between 
Mrk 509 and 3C273 $BeppoSAX$.}
\end{figure}

\begin{center}
\begin{table*}
\hspace{3.5cm}\hspace{3.0cm}\begin{minipage}{170mm}
\tabcolsep=3.0mm
\footnotesize
\caption{Broad band fits of SAX6. Column I: model number. Column II: photon index. Column III: relative normalization between the direct and reflected components. Column IV: energy centroid of the emission line. Column V: EW of the emission line. Column VI: energy centroid of the absorption line. Column VII: EW of the absorption line. Column VIII: $\chi$$^{2}$ and number of degrees of freedom. Errors are 90\% confidence for one interesting parameter.}
\scriptsize
\hspace{1.0cm}\begin{tabular}{c c c c c c c c }
\hline
\hline
&&&&&&&\\
Model & $\Gamma$& R & E$_{\rm em}$& EW$_{\rm em}$& E$_{\rm ab}$&EW$_{\rm ab}$&$\chi$$^{2}$/d.o.f\\
&&&&&&&\\
&&&keV & eV &keV &eV &\\
&&&&&&&\\
I&II&III&IV&V&VI&VII&VIII\\
&&&&&&&\\
\hline
&&&&&&&\\
1&1.61$^{+0.11}_{-0.05}$&0.86$^{+0.51}_{-0.60}$&6.41$^{+0.17}_{-0.14}$&143$^{+60}_{-61}$&&&110.9/119\\
&&&&&&&\\
2&1.58$^{+0.12}_{-0.05}$&0.99$^{+0.52}_{-0.58}$&6.37$^{+0.18}_{-0.20}$&128$^{+111}_{-53}$&5.60$^{+0.14}_{-0.12}$&93$^{+51}_{-46}$&98.5/117\\
&&&&&&&\\
\hline
\hline
\end{tabular}
\end{minipage}
\end{table*}
\end{center}

To fit the broad band spectrum, we followed De Rosa et 
al. (2004) who obtained  a better statistics by means of a simultaneous 
fit to the four 2000 pointings of Mrk 509. 
The first baseline model we tried consists of a flat primary power law 
($\Gamma$$\sim$1.6), plus a 
reflection component due to cold matter  (namely the PEXRAV model in 
XSPEC, Magdziarz \& Zdziarski 1995), and a superimposed FeK$\alpha$ 
emission line (table 2, model 1). 
At low energies (below $\sim$1.5 keV) the spectrum is dominated by the 
contribution of extra components. 
De Rosa et al. (2004) modeled this excess by 
adding a couple of black-body components.
To avoid the parameters of the fit to diverge, we fixed some of them
to the best fit values found by De Rosa et al. (2004): the temperatures
of the black bodies were set to be 71 and 240 eV, the high energy cut-off
was fixed to 83 keV, the inclination angle was fixed to 30$^{\circ}$ and 
abundances were fixed to solar values.

The fit improves significantly when a Gaussian absorption 
line is added to the model ($\Delta$$\chi$$^{2}$=12.4 for two more 
parameters, see model 2 in table 2), yielding  best fit parameters for the 
line: E$_{\rm ab}$=5.60$\pm$0.14 keV, and EW$_{\rm ab}$=93$\pm$50 eV. Best-fit 
unfolded spectrum and 
confidence contours for the absorption line parameters are shown in figure 3 
and 4, respectively.

\begin{figure}[htb!]
\psfig{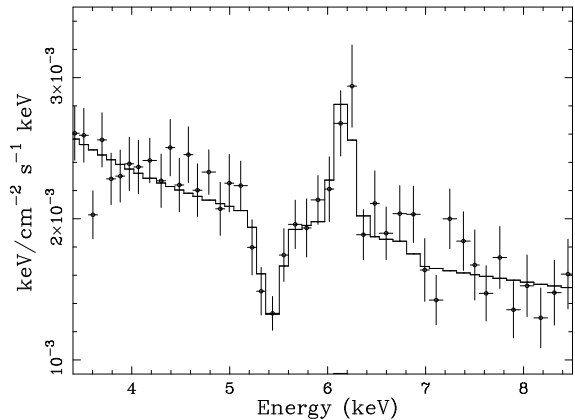}
\caption{Unfolded 3-10 keV spectrum measured in SAX6. The continuous line 
indicates model number 2 in Table 2. In producing the plot, the line widths 
were fixed to $\sigma$=0.1 keV for clarity.}
\end{figure}

\begin{figure}[htb!]
\psfig{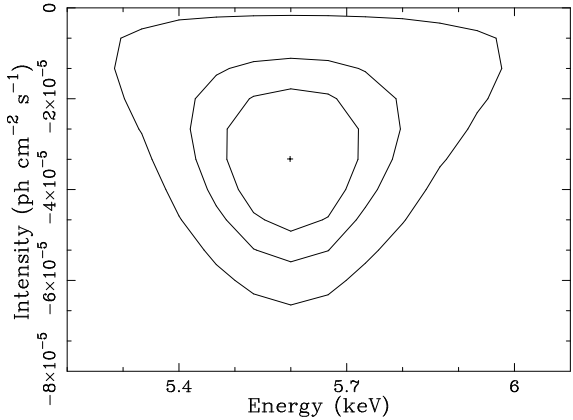}
\caption{99, 90 and 68\% confidence contours of the absorption line parameters detected in SAX6 (rest frame energy vs. intensity). The reference X-ray continuum used to calculate these contours is the number 2 in table  2.}
\end{figure}

We also tried a complex photoionized 
plasma model (namely the SIABS model in  XSPEC, Kinkhabwala et al. 2003) to 
fit the line.
Once the ionization state was fixed to FeXXVI, we obtained that 
the receding velocity of the putative absorbing matter is {\it v}$\sim$0.2c 
with a dispersion $\Delta${\it v}$\sim$3500 km s$^{-1}$, while an Iron 
absorbing column density of N$_{\rm Fe}$$\sim$1.3$\times$10$^{18}$ cm$^{-2}$. 
However, these parameters are only poorly constrained by the quality of the 
present data.

Following De Rosa et al. (2004) we tested also a ionized reflection model 
(Ross \& Fabian 1993). As expected, the available statistics prevents to 
firmly constrain all the parameters of the model. In this case, the 
improvement of the fit corresponds to a $\Delta$$\chi$$^{2}$=12.1.
The best fit value  of the 
ionization parameter is log$\xi$$\sim$3.5 ($\xi=4\pi F_x/n_{\rm H}$, where 
F$_x$ is the 0.01-100 keV flux illuminating a slab of gas with solar 
abundance and constant hydrogen number density
n$_{\rm H}$=10$^{15}$ cm$^{-3}$. The incident flux is assumed to be a power law
 with a sharp high energy cut-off at E$\sim$100 keV).

We searched for absorption lines between 
5 and 6 keV also in the other $BeppoSAX$ and $XMM$-$Newton$ observations but 
obtained  only upper-limits ranging between 10 and 50 eV 
(at 90\% confidence).

\subsection{SAX6 observation: time-resolved analysis}

The occurrence of such an absorption feature in only one out of six $BeppoSAX$ 
observations of Mrk 509 suggests a sporadic nature. 
This led us to dissect the SAX6 observation so as to further investigate  
the feature variability. We found a best trade-off between time 
resolution and collected photons (available statistics) in sectioning the 
observation in five intervals represented by the time bins of the light curves 
shown in figure 5, where the light curves in the 5-5.8 keV band (upper panel) 
and in the 2-10 keV band (lower panel) are presented.

\begin{figure}[htb!]
\psfig{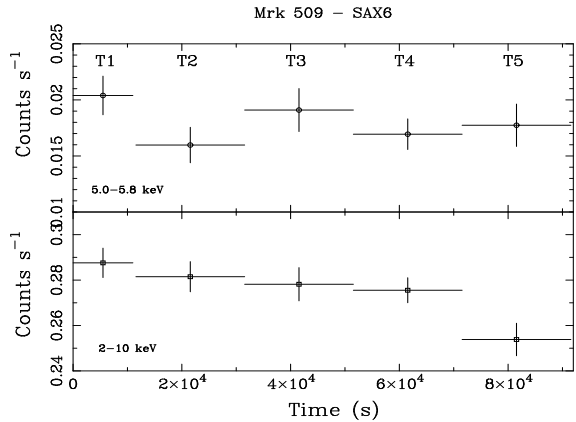}
\caption{SAX6 light curves in the 5-5.8 keV range (upper panel) and in the 
2-10 keV range (lower panel).}
\end{figure}

\begin{figure}
\psfig{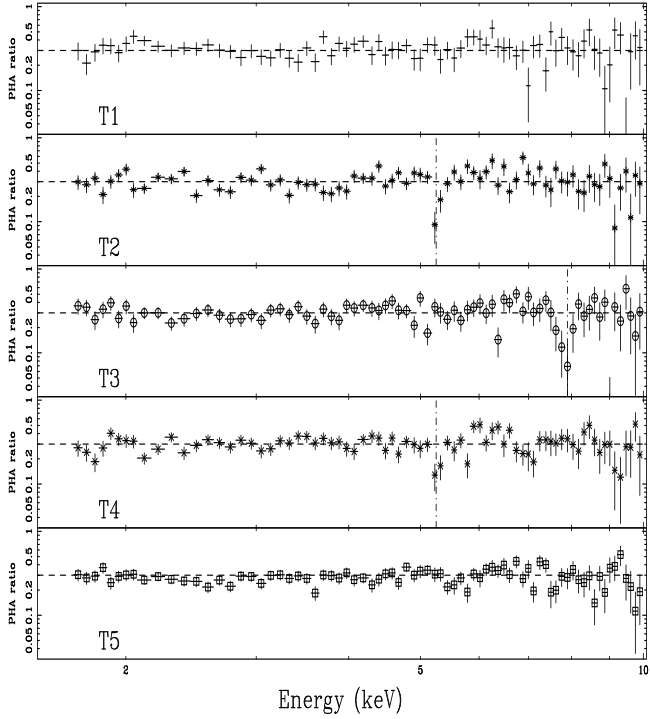}
\caption{Mrk 509/3C273 PHA ratios obtained dividing the SAX6 observation 
in 5 intervals (see text for details).}
\end{figure}

To exclude systematic effects, we performed the 
PHA ratios with 3C273 data also in these time intervals (Figure 6). During 
periods T2, T3 and T4, three different  ``absorption features'' are clearly 
visible at E $\sim$5.3 keV (T2 and T4) and 7.9 keV (T3) (observer frame). 
The count drops account for deviations as large as $\sim$60\% and 40\% in
T2 and T4, respectively. 

\begin{center}
\begin{figure}
\hspace{0.5cm}\psfig{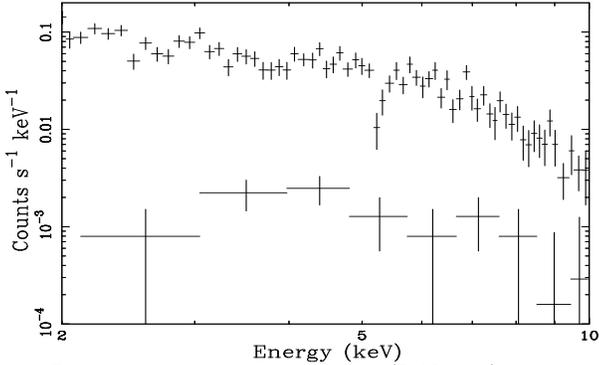}
\caption{Source and local background (2-10 keV) spectra during T2 which 
clearly show how, at the energies of the absorption line, the background 
contributes to $\sim$5\% of the source counts. Similar results are 
obtained for T3 and T4 periods.}
\end{figure} 
\end{center}

\vspace{-1.0cm}

A check of the relative importance and shape of the background during these 
short intervals (see figure 7 for the T2 interval) 
confirmed that no anomalous features were present in the local 
background during T2, T3 and T4 periods.

To model these spectral features we restricted the spectral analysis only 
to the $\sim$2-10 keV MECS data. 
The data were fitted with a simple power-law model.  
We find that the addition of absorption features increases
significantly the quality of the fit in T2, T3 and T4 (see table 3). 
The addition of a 
narrow or broad FeK$_{\alpha}$ emission line is not required by the data
but we have checked and found that the absorption line detections are also 
robust to the inclusion of either narrow or broad emission lines above and/or 
below the absorption line energy. 

Figure 8 shows the confidence contours plots 
for the (rest frame) parameters of interest of these absorption lines.
The line widths are consistent with the instrumental energy resolution.
We also tried to fit the feature at $\sim$8 keV during T3 with an 
absorption edge but the result was a poorer fit 
($\Delta$$\chi$$^2$$\sim$14 for the same degrees of 
freedom) when compared with that obtained with the absorption Gaussian line.

\begin{center}
\begin{table*}
\caption{Time resolved analysis of SAX6. Column I: name of the time interval. Column II: energy centroid of the absorption line. Column III: line intensity. Column IV: line EW. Column V: $\chi$$^{2}$ variation after inclusion of an absorption Gaussian line to a simple power law model. Column VI: F-test significance of the absorption line.  Errors are 90\% confidence for one interesting parameter.}
\tabcolsep=3.0mm
\footnotesize
\scriptsize
\hspace{3.3cm}
\begin{tabular}{l c c c c }
\hline
  &  & & &   \\
SAX6 Period  & E & Line Intensity & EW$_{\rm ab}$ &$\Delta\chi^{2}$ \\
  &  & & &  \\
  & (keV) &(10$^{-5}$ph s$^{-1}$cm$^{-2}$) & eV & \\
  &  & & &  \\
I  & II & III & IV & V \\
  &  & & &  \\
\hline
   &  & &   & \\
T2&5.50$^{+0.12}_{-0.12}$& $-$7.82$^{+3.25}_{-3.49}$&195$^{+81}_{-85}$&16.8\\
   &  & &  &  \\
T3&8.14$^{+0.15}_{-0.13}$& $-$8.48$^{+4.14}_{-4.49}$&383$^{+96}_{-203}$&16.4\\
   &  & & &  \\
T4&5.45$^{+0.15}_{-0.15}$& $-$6.97$^{+3.65}_{-3.66}$&173$^{+90}_{-90}$ &9.6\\
   &  & &  &\\
\hline
\hline
\end{tabular}
\end{table*}
\end{center}


\begin{figure}[htb!]
\psfig{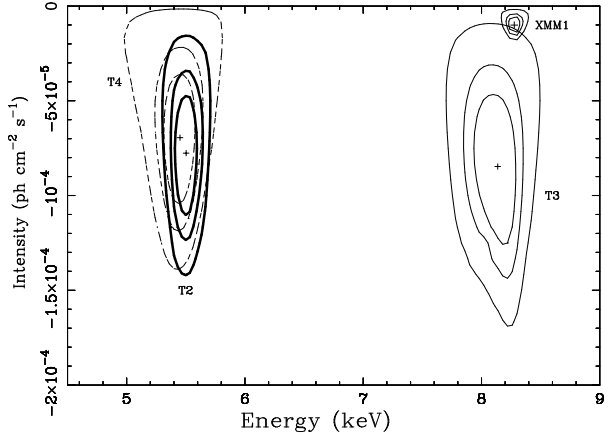}
\caption{Confidence contours for parameters of interest of the absorption 
features measured during periods T2 (thick solid line), T3 (solid line) 
and T4 (dashed line) of SAX6 and XMM1. The line width was fixed to 
$\sigma$=0 eV. Rest-frame line energies are reported.}
\end{figure}

We searched for variability in the absorption lines 
by fixing their energy at 5.5 and 8.1 keV, and computing the corresponding 
intensities during T1 to T5 with respect to a simple power law model.
The 90\% upper limit on EW$_{\rm ab}$ during T5 
is not consistent with what measured in T2 (panel (a) of figure 9) and only 
marginally with T4. 
On the other hand, the $\sim$8.1 keV feature detected in T3 is well above the 
90\% upper limits obtained in the other time intervals (panel (b) of figure 9).

\begin{figure}[htb!]
\psfig{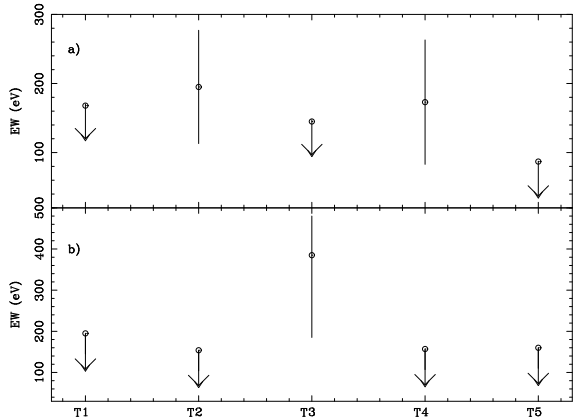}
\caption{SAX6 time resolved analysis. Panel a: 5.5 keV absorption feature 
EW$_{\rm ab}$ vs. time intervals. The EW$_{\rm ab}$ obtained 
during T1, T3 and T5 have been obtained fixing the absorption line energy 
at 5.5 keV (rest-frame). The continuum is fitted by a simple power-law model.
Panel b: Same as panel a) with line energy fixed at 8.1 keV (rest-frame).
Error bars and upper limits are at 90\% confidence level.}
\end{figure}

\vspace{-0.5cm}
\subsection {{\it XMM-Newton} observation of Oct. 25, 2000 (XMM1)}

This observation  was previously analyzed by Page, Davis 
\& Salvi (2003). These authors commented the presence of residuals at 
E$\sim$8 keV to their best-fit model, but did not attempt to model them. 

Fitting the 2-10 keV data with a simple power-law, the addition of a 
narrow absorption line at E=8.27$\pm$0.05 keV with an 
EW$_{\rm ab}$=40$\pm$20 eV, leads to a reduction of the $\Delta$$\chi$$^{2}$=12.5 (figure 8). 
As for SAX6, also in this case the modeling
of this absorption feature with an edge leads to a worse 
fit ($\Delta$$\chi$$^2$$\sim$10 for the same number of degrees of freedom).


In this energy range, the background of the 
EPIC CCDs is strongly affected by a Cu activated emission line. The 
net effect of an inadequate removal of this background feature could, in 
principle, cause the presence of an absorption and/or emission feature in the 
source spectrum. 
We tested this possibility by using different background regions. This did not 
change significantly the results 
on the absorption line parameters Moreover, at the energy of the absorption
feature, the background contributes for $\lesssim$5\% to the source 
counts (figure 10), thus excluding that the $\sim$8 keV 
absorption feature is due to background subtraction problems. 

\begin{figure}[!]
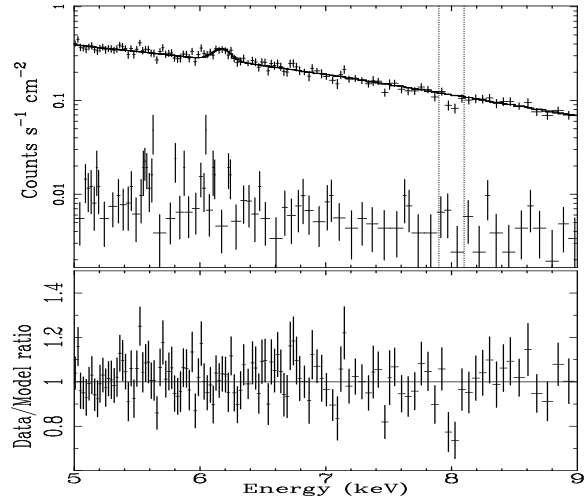

\psfig{figure=fig_10.ps,height=3.5cm,width=7.5cm,angle=270}
\hspace{-0.042cm}\psfig{figure=ratio_fig10.ps,height=3.cm,width=7.62cm,angle=270}
\caption{Upper panel: source and background (5-9 keV) spectra during XMM1. 
Dotted lines highlight the energy range of the detected 
absorption line. A FeK$\alpha$ narrow emission line has been added to the
model. Lower panel: data-to-model ratio with respect to a spectral model
composed by a power-law plus an narrow emission line.}
\end{figure}

\subsection{Other $BeppoSAX$ and {\it XMM-Newton} Observations of Mrk 509}

We performed time resolved search for absorption features also in the other 
observations of Mrk 509 performed by $BeppoSAX$ and {\it XMM-Newton}, but 
without success. This confirms the indication that
these features are of transient nature and partially explains why 
these features were not detected before despite that the same data were 
already analyzed. Moreover, the occurrence of these features seems
to be independent from the source state. 
So a bias against their detection is expected since they are expected to be 
washed out in spectra averaged over long exposures.

\subsection{On the significance of the absorption lines}

A word of caution should be stated for the statistical significance 
of the absorption lines detected in SAX6 and 
XMM1 at $\sim$5.5 and $\sim$8.2 keV, respectively. If the F-test is applied
to these observations, the output is that these features are significant
to a $\sim$99.9\% confidence level. Nonetheless this does not account for the 
total number of trials  performed, as well as for the total number of 
resolution elements used for the line detection (see, e.g., Protassov et al. 
2002; Porquet et al. 2004; Yaqoob \& Serlemitsos 2005). If we consider
only the 5.5 keV feature and the average spectra, we find that the 
F-test significance decreases to $\sim$99.2\% because of the number of 
trials involved (the eight $BeppoSAX$ and $XMM$-$Newton$ observations).
The same arguments could be applied for the time-resolved analysis (a total
of 33 time resolved spectra) thus reducing the overall significance
in detecting a single absorption line at $\sim$96.7\%.
Moreover, Protassov et al. (2002) discussed how the F-test like indicators 
could fail in assessing the real significance of an absorption feature maybe
overestimating it. We cannot, therefore, strongly fix the ``true'' 
significance of our results that could be, most probably, lower than what 
deduced by the application of the F-test. 

It is encouraging, however, that the four lines detected with 
the time resolved analysis of the data, have been found to be consistent
with only two energies. Moreover, recently, detections of both blue- and 
red-shifted lines have been claimed by some authors 
(Pounds et al. 2003a,b, Yaqoob \& Serlemitsos 2005, Matt et al. 2005), thus 
suggesting that the features 
reported here are not a statistical artifact. To conclude, it is worth noting 
here that confirmations by means of instruments with larger collecting area 
and/or greater energy resolutions, are needed to definitively fix the 
veridicity of the spectral and time patterns measured here.


\section{Discussion}

Evidences for absorption features in the X-ray spectrum
of Mrk 509 have been detected at $\sim$5.5 keV (twice in BeppoSAX data) and 
at $\sim$8.2 keV (in both $BeppoSAX$ and $XMM$-$Newton$ non-simultaneous data).
Both features appear to be transient on time scales that could be as short
as $\sim$20 ks.

As previously proposed for other Seyferts and QSOs, the most natural 
explanation of such narrow components is in terms
of both blue-shifted and red-shifted resonant absorption lines 
from H-like or He-like Iron (Nandra et al 1999; 
Pounds et al. 2003; Reeves, O'Brien \& Ward 2003; Longinotti et al. 
2003; Yaqoob \& Serlemitsos 2005; Matt et al. 2005).
This is also consistent with the fact that 
the best-fit of the spectrum requires 
an ionized reflection component with log$\xi$$\sim$3.5, 
thus implying the presence of FeXXV and FeXXVI (De Rosa et al. 2004).

If associated to FeXXVI (6.96 keV rest-frame), the inferred shifts in 
energy correspond to receding velocities of (0.21$\pm$0.02)c (T2 and T4) and 
to approaching velocities of (0.16$\pm$0.02)c (T3) and 
(0.19$\pm$0.02)c (XMM1). If associated mainly to FeXXV (6.7 keV rest-frame), 
the velocities become $\sim$0.18, 0.20 and 0.23c, respectively. 

Nandra et al. (1999) first discussed the interesting possibility 
that the red-shifted line seen in NGC 3516 could be the direct signature 
of matter inflow/infall. 
Alternatively, Ruszkowski \& Fabian (2000) explained how the NGC 3516 feature 
could be due to a warm plasma surrounding the X-ray source. In their picture, 
the absorbing matter is located close to the black hole and the resonant 
absorption line is red-shifted by the gravitational field produced by the 
SMBH. A similar interpretation has been given by Yaqoob \& Serlemitsos (2005) 
for the red-shifted Iron line observed with the Chandra HETG in the QSO 
E1821+643. Present results cannot rule out that the line red-shifts 
measured in T2 and T4 are partially or even totally due to gravitational 
red-shift. 
Nonetheless, the data seem to require either a 
transient or an ``un-steady'' absorber in order to 
explain the sporadic presence of the absorption features. 

The temporal succession of the 
absorption events measured in T2-T3-T4 (red-blue-red shifts) seems to indicate 
a physical relation between the red and blue-shifted components. 
The present data, in fact, suggest a picture in which 
we are observing the succession of matter infalls and ejecta that 
sporadically obscure the X-ray continuum source, i.e. indicating
the presence of non-circular motions close to the SMBH. The use of
the SIABS model (see \S 2.1, Kinkhabwala et al. 2003) to fit the entire 
SAX6 data gives an Iron column density
N$_{\rm Fe}$$\sim$1.3$\times$10$^{18}$ cm$^{-2}$ for the red-shifted lines, 
which, assuming 
solar abundances, corresponds to a total column of 
N$_{\rm H}$$\sim$6$\times$10$^{22}$ cm$^{-2}$. Considering the EW$_{\rm ab}$ 
obtained in shorter periods and using the ``curves of growth'' presented 
in Kotani et al. (2000), we obtained the absorbing
columns: N$_{\rm H}$$\sim$1.3$\times$10$^{23}$ cm$^{-2}$ in T2-T3-T4 and  
N$_{\rm H}$$\sim$4$\times$10$^{22}$ cm$^{-2}$ in XMM1.  
Assuming 20 ks (i.e. the duration of T2, T3, and T4) as 
the characteristic variability timescale, the upper limit to the dimensions
 of the absorbing regions can be estimated to be $d$$\sim$6$\times$10$^{14}$ 
cm.
This leads to an upper limit on the mass involved of 
$\sim$0.3-1.2$\times$10$^{28}$ g ($\sim$1.5-6$\times$10$^{-5}$M$_\odot$), depending on the geometry of the absorber. 
The related density lower limit is, thus, $\sim$0.5-2$\times$10$^{8}$ atoms 
cm$^{-3}$. 
From the measured velocity $v$$\sim$0.2c it is then possible to infer the 
kinetic energy of the moving matter 
E$_{kin}$$\leq$$\Gamma$Mc=9$\times$10$^{47}$ erg, 
where $\Gamma$ is the bulk Lorentz factor and M is the absorber mass. 
The difference 
between the column densities measured in SAX6 and XMM1 indicates that
either the density or the geometry or the ionization state of the intervening 
matter has varied between the two epochs ($\sim$one month apart).

Our results are in good agreement with what predicted by the ``aborted 
jet'' model for the production of X-rays in radio quiet AGN (Ghisellini, 
Matt \& Haardt 2004). In this framework, X-rays are produced via the 
extraction of rotational energy from the BH by repeated expulsions of 
blobs/clouds of matter with initial speed below the escape velocity. The 
collision between blobs determine the transmission of a significant amount of 
the kinetic energy to the gas contained in the blobs. This originates 
a hot plasma which comptonizes to X-rays the seed UV photons from the accretion
disk. At the end of the process, about 5-10\% of the initial kinetic power
is transformed into X-rays that dominate the nuclear emission when the
source is in high luminosity states. 
From our data we can estimate the jet power ignition to be 
$\sim$E$_{kin}$/t$_{obs}$ ($\leq$10$^{44}$ erg s$^{-1}$), where t$_{obs}$ is 
the total length of the SAX6 observation. Thus, during SAX6 and XMM1, the 
``aborted jet'' contributes for less than 5-10\% to the total X-ray luminosity 
observed. 
This value is in agreement with the theory  since 
it is expected that the jet contribution to the X-ray 
emission is negligible during the low luminosity states like SAX6 and XMM1. 
In fact, these are predicted to be dominated by the disk-corona 
X-ray emission.

Finally, our results are qualitatively in agreement also with more traditional 
models. Magneto-hydrodynamic (MHD) simulations of the 
matter inflow onto a black hole via an 
accretion disk (De Villiers, Hawley, \& Krolik 2003) predict that the 
accretion flow of the disk is almost stable, but, close to the SMBH, local 
instabilities may lead to the formation 
of infalling blobs/winds. Coupled with the accretion flow, these  
simulations also predict outflows of warm/hot matter 
that supply the corona with both gas and magnetic energy, and 
lead to the formation of high velocity ejecta.
Similarly, what observed in Mrk 509 
may be interpreted in the view of the ``Thundercloud model'' 
(Merloni \& Fabian 2001) that predicts turbulent motions of gas in a 
``patchy corona'' over the accretion disk. In this framework, corona/disk 
instabilities could produce episodes of matter ejecta and infalls.

\section{Conclusions} 

Understanding gas inflows/outflows in AGNs is to get insight into 
accretion/expulsion processes, two of the most fundamental questions 
regarding black hole systems. To date, data from several Seyfert galaxies and 
QSOs show signatures of narrow, blue-shifted absorption lines 
explained as due to Iron resonant absorption in outflowing gas. 
In this work, we report evidence of both red-shifted and 
blue-shifted transient absorption lines in the X-ray spectrum of the Seyfert 
galaxy Mrk 509.
If confirmed, the red-shifted lines would be particularly interesting 
because they could be one of the best direct evidence of matter free-falling 
onto a SMBH.  
Broadly speaking, the data presented here are in good
agreement with most theoretical models which imply important radial 
motions close to the SMBH (such as, for example, the ``aborted jet'' model).
Other scenarios which explain the line red-shifts in terms of 
gravitational red-shift cannot be ruled out but seem to require a 
higher degree of complexity to account for the pattern and 
sporadic nature of the absorption lines. 
A detailed analysis of the line 
profile is needed in order to test these hypotheses.
If confirmed, these results may offer a new potential to study the kinematics 
and dynamics of the gas close to the SMBH via detailed absorption line 
X-ray spectroscopy. If the inflow is due to a wind rather than blobs, 
one would expect the features to exhibit a typical inverted P-Cygni 
profile (Edwards et al. 1994). On the other hand, a blob scenario, as 
favored here, would provide test-particles suitable to verify predictions 
of General Relativity in strong gravitational fields.

\acknowledgements  

We are very grateful to G.G.C Palumbo, P. Grandi, G. Ghisellini and G. Matt 
for useful discussions. We thank the anonymous referee for her/his constructive
comments.
This research made use of the ASI Science Data Center and {\it XMM-Newton} 
Science Operation Center databases. 
M.D. and A.D.R gratefully acknowledge financial support from 
the Italian Space Agency (ASI).

\end{document}